\documentclass{appolb}
\usepackage{graphicx}
\usepackage{amsmath}
\usepackage{wasysym}
\usepackage{amsfonts}
\usepackage{amssymb}
\usepackage{amsmath}
\usepackage{pifont}
\usepackage[varg]{txfonts}
\usepackage{cite}
\newcommand{\nn}{\nonumber}
\newcommand{\mbo}[1]{$ #1 $}

\newcommand{\tv}{\mbox{TeV}}
\newcommand{\epo}{\;.}

\newcommand{\I}{\mathrm{i}}

\newcommand{\be}{\begin{equation}}
\newcommand{\ee}{\end{equation}}
\newcommand{\bea}{\begin{eqnarray}}
\newcommand{\eea}{\end{eqnarray}}

\newcommand{\epm}{e^+e^-}
\newcommand{\amuexp}{a_\mu^{\mathrm{exp}}}

\newcommand{\power}[1]{\times 10^{#1}}

\def\pipi{\pi^+ \pi^-}
\def\amu{a_\mu}
\def\amuh{a_\mu^{{\mathrm{HVP,\ LO}}}}
\def\amuLbL{a_\mu^{{\mathrm{LbL,\ LO}}}}
\begin{document}
\title{%
\vskip-3cm{\baselineskip14pt
\centerline{\small HU-EP-26/03\hfill January 2026}}
\vskip1.5cm
Lepton Magnetic Moments: What They Tell Us%
\thanks{Presented at \textit{Matter to the Deepest} Conference 2025 Katowice/Poland September 15-19.}}%
\author{Fred Jegerlehner
\address{Humboldt-Universit\"at zu Berlin, Institut f\"ur Physik, Newtonstrasse 15,
D-12489 Berlin, Germany}
}
\maketitle
\thispagestyle{empty}
\begin{abstract}
 Recently, the exciting new Fermilab (FNAL) Muon g-2 measurement
 impressively confirmed the final Brookhaven (BNL) result from 2004
 and, with a result four times more precise, has launched a new
 serious attack on the Standard Model (SM). On the theoretical side,
 ab initio lattice QCD (LQCD) calculations of hadronic vacuum polarization
 have made remarkable progress. They are now the new standard for
 studying the leading non-perturbative contributions, which have
 previously hindered matching with the precision required for full
 exploitation of the experimental results. The lattice results
 affected both leading hadronic contributions -- the hadronic vacuum
 polarization (HVP) and the hadronic light-by-light (HLbL)
 contributions -- by increasing the previously generally accepted
 $e^+e^- \to$ hadrons based dispersion relation results. The shifts
 reduced the discrepancy between theory and experiment to “nothing
 missing.” One of the most prominent signs of ``Beyond the Standard
 Model'' (BSM) physics has disappeared: the SM appears validated more
 than ever, in agreement with what also searches at the Large Hadron
 Collider (LHC) at CERN tell us! A triumph of the SM, in spite of the
 fact that the SM cannot explain known cosmological puzzles like dark
 matter or baryogenesis and why neutrino masses are so tiny, the
 absence of strong CP violation for example. I also argue that the
 discrepancy between the data-driven dispersive result and the lattice
 QCD results for the hadronic vacuum polarization can be largely
 explained by correcting the $\epm$ data for $\rho^0-\gamma$ mixing
 effects.
\end{abstract}
\section{Introduction}
This text is an update of an earlier
article~\cite{Jegerlehner:2018zrj} and the more detailed
book~\cite{Jegerlehner:2017gek}, which were written when the 2004 BNL
measurement of $a_\mu= (g_\mu-2)/2 $ at 540 ppb~\cite{Muong-2:2006rrc}
had set the precision limit before the Fermilab Muon g-2
experiment~\cite{Muong-2:2021ojo,Muong-2:2023cdq} provided new, more
precise measurements of $a_\mu$, which recently reached the milestone
of 124 ppb~\cite{Muong-2:2023cdq}. With four times the precision and a
statistical error roughly equal to the combined systematic error, the
SM was subjected to an unprecedented precision
test. While the Fermilab experiment confirmed the final BNL result,
the theoretical prediction moved closer to the experimental result due
to significant advances in ab initio lattice QCD calculations of the
non-perturbative QCD contributions that had been an obstacle to the
precision of the SM prediction of $a_\mu$.

The story of the lepton magnetic moments is the story of ``The closer
you look the more there is to see'' and highlights the SM to the
deepest. The magnetic moments of leptons (\mbo{g_\ell} factors) and
the associated anomalies of the muon and
electron play a decisive role in high-precision verification of the
Standard Model. These quantities are not only among the most precisely
measured in particle physics, but can also be predicted with high
accuracy. At the same time, they offer promising insights into physics
beyond the Standard Model. The key relationship for testing possible
new heavy states is\\[-6mm]
$$\frac{\delta a_\ell}{a_\ell} \propto \frac{m^2_\ell}{M^2} ~~~(M \gg
m_\ell),$$ where $m_\ell$ is the mass of the lepton and $M$ is the
mass scale of new physics. This equation shows that the fractional
change in $a_\ell$ is proportional to the square of the lepton mass
over the square of the new physics scale, making the muon about
$(m_\mu/m_e)^2 \sim 4\times10^4$ times more sensitive to BSM physics
than the electron. For the muon, theory and experiment now agree to
{\sf 9 decimal places}, so that $\delta a_\mu \sim \frac{\alpha}{\pi}
\frac{m_\mu^2}{M^2_{\rm NP}}$ can effectively probe intermediate mass
scales beyond the heaviest SM ingredient, the top quark mass \mbo{M_t}
up to about \mbo{1~\tv}, a region where new heavy states are
essentially ruled out directly by the LHC.

In any case, the ``Muon-g-2 drama'' will have an essential impact on
which SM extensions remain viable. Paradigms such as naturalness and
the supposed related hierarchy problem are definitely not constructive, and
new guiding principles must take their place. The most fruitful
approach here is to search for emergent structures, as outlined
in~\cite{Jegerlehner:2021vqz,Bass:2025eho}, for example.
\section{A new milestone in the experimental determination of the muon anomaly}
\begin{figure}
\centering
\includegraphics[width=0.46\textwidth]{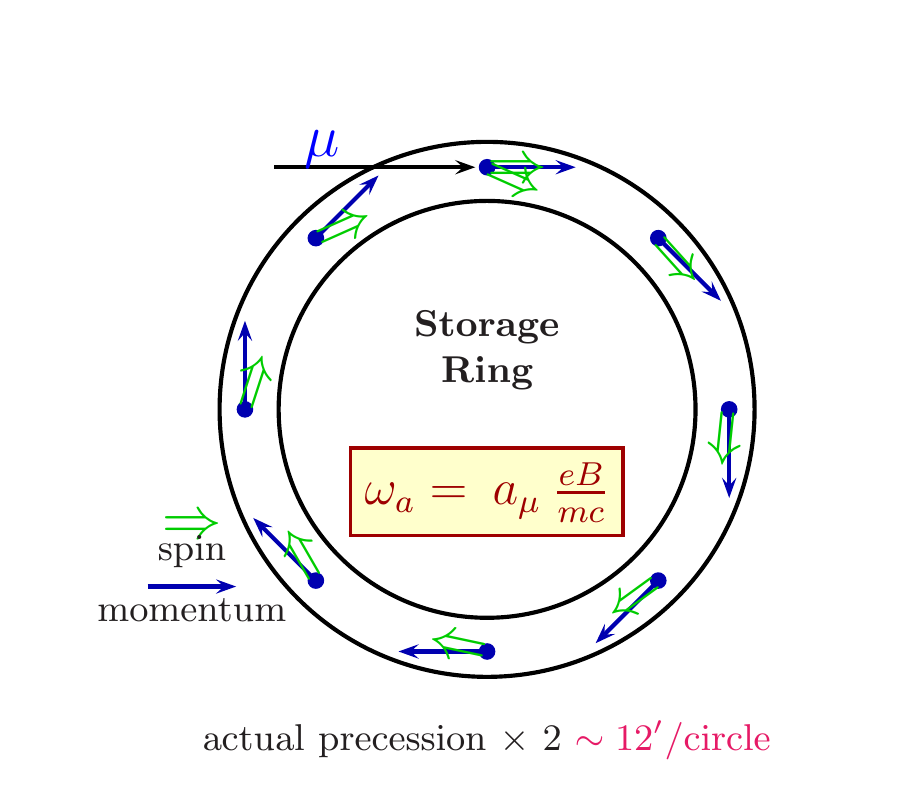}
\includegraphics[width=0.50\textwidth]{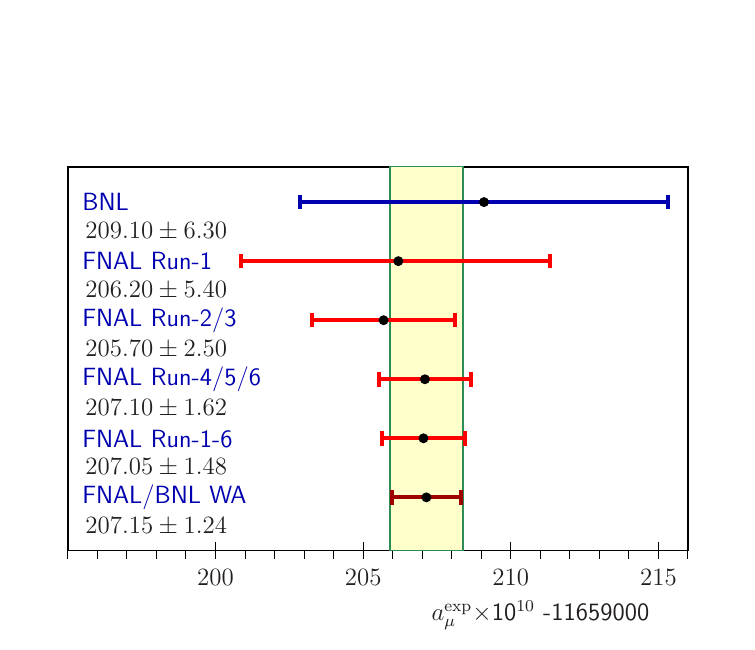}
\caption{Left: At the Magic Energy (muon beam energy $\sim$ 3.1~GeV),
  the angular frequency \mbo{\vec{\omega_a}} is directly proportional
  to the magnetic field \mbo{\vec{B}}, so that the $a_\mu$ measurement
  is a frequency counting experiment. Right: The measurements of
  $a_\mu$ by the Muon g-2 Collaboration at the Fermi National
  Accelerator Laboratory (FNAL) have significantly improved on the
  results obtained by Brookhaven National Laboratory (BNL) in 2004.}
\label{fig:wiggles} 
\end{figure}

Any particle with spin \mbo{\vec{s}\,} has a magnetic moment \mbo{\vec{\mu}}
(internal current circulating)
$$
\vec{\mu}={ g_\mu}\:\frac{e \hbar}{2m_\mu c}\:\vec{s}\;\; ;\;\;\;
{g_\mu=2\:(1+{a_\mu})}
$$
The extreme precision one is able to reach in measuring and
calculating the lepton anomalous magnetic moments is possible by
taking the simplest object you can think of in the static limit which
is provided with the electromagnetic vertex\\[-8mm]
\begin{figure}[h!]
\hspace*{12mm}\includegraphics[height=2cm]{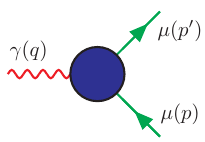}

\vspace*{-1.3cm}

\hspace{4cm}$ =(-\I e)\:\bar{u}(p')\left[\gamma^\mu F_1(q^2)+\I\,\frac{\sigma^{\mu\nu}q_\nu}{2m_\mu}F_2(q^2)\right]u(p)$
\end{figure}

\noindent where \be F_1(0)=1\;\;;\;\;\; F_2(0)=a_\mu \ee The point is that
$a_\mu$ is responsible for the Larmor (spin) precession frequency
\mbo{\vec{\omega}_a} which shows up when polarized muons are orbiting in
a homogeneous magnetic field as illustrated in Fig.~\ref{fig:wiggles}.
This\footnote{ $\gamma$ is the Lorentz factor of relativistic time
dilation. It is adjusted so that $a_\mu-1/(\gamma^2-1)=0$ to cancel
out the dependence on the electric field $\vec{E}$ needed to focus the
beam.} ``magic $\gamma$'' muon storage ring principle developed at
CERN has been applied in later experiments at Brookhaven (BNL) and
Fermilab (FNAL)\footnote{There are good reasons why a very different
experimental technique for measuring $a_\mu$ is much-needed. To come
-- novel complementary experiment at J-PARC in
Japan~\cite{Saito:2012zz,Mibe:2011zz,Venanzoni:2025rsd} where a strict
{$\vec{E}=0$} cavity is utilized. In place of the ultra relativistic
(CERN, BNL, Fermilab)) muons one then can work with ultra cold
(J-PARC) muons such that one obtains a measurement with very different
systematics.}. The new Fermilab Muon $g-2$ measurement (see Fig.~\ref{fig:wiggles}), applying
sophisticated improved technology, brightly confirmed the 2004 final
BNL result\\[-4mm]
\begin{equation}
  a_\mu[\rm BNL]= 11659209.1(5.4)(3.3)[6.3]\power{-10}
\end{equation}
now improved to\\[-4mm]
\begin{equation}
a_\mu[{\rm FNAL}]= 11659207.05(1.14)(0.95)[1.48]\power{-10}
\end{equation}
and potentially triggers a new serious attack on
the SM. The combined result is
\begin{equation}
\amuexp=(11\,659\,207.15\pm 0.80\pm 0.95)[1.24])\power{-10}
     ~~~ (\mbox{FNAL/BNL})\epo
\label{amuexpWA}
\end{equation}
This is a remarkable achievement with the precision of 124 ppb today
from 540 ppb at BNL, a factor 4 improvement with statistical error
as small as the systematic one.
  
\section{Theory Today}
The theoretical prediction of $\amu$ includes the entire SM, and the
result presented in the White Paper in 2020 (WP20) reads~\cite{Aoyama:2020ynm}:\footnote{ Recently, in 2025, the WP20 result
was updated~\cite{Aliberti:2025beg}, which we will refer to below as
WP25.}
\begin{eqnarray}
a_\mu^{\rm SM} =F_2(0) &=& a_\mu^{\left({
    \rm QED}+{\rm EW}+{\rm HVP}_{\rm LO}+{\rm HVP}_{\rm
    NLO}+{\rm HVP}_{\rm NNLO}+{\rm HLbL}_ {\rm LO}+{\rm HLbL}_{\rm
    NLO}\right)} \nn \\ &=&0.00116591810(43)\epo
\end{eqnarray}
This result, worked out by the Muon g-2 Theory
Initiative, relies on the values
\mbo{\amuh=693.1(4.0)\power{-10}} and \mbo{\amuLbL=90(17)\power{-10}}
for the problematic leading hadronic contributions\footnote{My value
from 2007 was $ a_\mu^{\rm SM}=0.00116591793 (68)$
in~\cite{Jegerlehner:2007xe,Jegerlehner:2008zza}, i.e., the central
value of WP20 changed only by $81.0-79.3=1.7$ in $10^{-10}$, while the
error could be reduced by 30\% mainly due to more precise $\epm$ data
in the calculation of the hadronic contributions. The WP20 result for
$\amuh$ was obtained by the $\epm$-data driven dispersion relation
approach. The relevant $\epm$ data have been obtained from the energy
scan data by CMD-2~\cite{CMD-2:2001ski}, SND~\cite{Achasov:2006vp},
SND20~\cite{SND:2020nwa} and CMD-3~\cite{CMD-3:2023alj} and the
Initial State Radiation (ISR) data from
KLOE~\cite{KLOE:2004lnj,KLOE:2008fmq,KLOE:2012anl,KLOE-2:2017fda},
BaBar~\cite{BaBar:2009wpw,BaBar:2012bdw}, BESIII~\cite{BESIII:2015equ}
and CLEO-c~\cite{Xiao:2017dqv}}. An interesting point concerns the
inclusion of \mbo{\tau} decay-spectra\footnote{This was pioneered by
Davier et al. in ~\cite{Alemany:1997tn}.} in the calculation of
$\amuh$, which leads to the increased value $\amuh [e^+e^-\, +\,
  \tau]=705.3\pm 4.5\power{-10}$~\cite{Davier:2010fmf}. Similarly,
calculations of $\amuh$ based solely on $\tau$ data\footnote{$\tau^-
\to \nu_\tau \pi^0 \pi^-$ spectra were recorded by
ALEPH~\cite{ALEPH:1997fek,ALEPH:2005qgp}, OPAL~\cite{OPAL:1998rrm},
CLEO~\cite{CLEO:1999dln} and Belle~\cite{Belle:2008xpe}.}~\cite{Masjuan:2023qsp,Castro:2024prg} yield $\amuh
[\tau]=703.1^{+4.1}_{-4.0} \times 10^{-10}$. We learn that by
including the $\tau$ data, both calculations yield values close to the
known lattice QCD result from the Budapest-Marseille-Wuppertal (BMW)
Collaboration~\cite{Borsanyi:2020mff}: $\amuh [\rm LQCD: BMW]=707.5\pm
5.5 \times 10^{-10}$. These results compare to the value \mbo{\amuh
  =717.0 \pm 1.5 \times 10^{-10}} which would close the mismatch
between theory and experiment. Also, if we compare the result
determined from $\tau$ data by Belle 2008: $ a_\mu^{\pi\pi}[2
  m_\pi, 1.8{\rm GeV}]=(523.5\pm3.9)\times 10^{-10}\: (\tau:\,{\rm
  Belle})\,,$ with the result obtained from the $e^+e^-$ data $
a_\mu^{\pi\pi}[2 m_\pi, 1.8{\rm GeV}]=(504.6\pm10.1,1)\times
10^{-10}\: (e^+e^-:\,\mbox{CMD-2,\,SND})\,,$ there is a difference of
$18.9 \times 10^{-10}$. Adding this difference to the dispersive
result of the $\epm$ data $\amuh [e^+e^-] = (694.79 \pm 4.18)\,\times
10^{-10}$, we obtain $713.7 \times10^{-10}$. Recently, the
energy-scan experiment CMD-3 reported an $e^+e^- \to
\pipi$ cross-section measurement in the energy range from 0.327 to 1.2
GeV, which also yielded the significantly higher value
$\amuh(2\pi,\,\mbox{CMD-3})= (526.0\pm4.2)\,\times 10^{-10}$, which,
compared to the corresponding WP20 estimate
$(506.0\pm 3.4)\,\times 10^{-10}|_{\rm WP20}$,
results in a difference of $ 20.0 \times 10^{-10}$.
Adding this to $\amuh[e^+e^-] = (693.1\pm 4.0)\power{-10}|_{\rm WP}$
yields $ (713.1\pm 4.7)\power{-10}$, a
surprising result since the second experiment with the SND detector at
the same storage ring yielded a result of $\amuh(2\pi,\, { \rm
  SND20})=(508.3\pm4.2)\,\times 10^{-10}$, which roughly corresponds
to the other $\epm$-data-based analyses. The large difference between
the CMD-3 value and the SND20 result (both $\epm$ scan results from
the $\epm$ collider facility in Novosibirsk) is another
puzzle\footnote{CMD-3 seems to apply different radiative corrections
(sQED vs RLA?) from other experiments;
see~\cite{Ignatov:2022iou,CMD-3:2023alj}.}!

While the $\epm \to \pipi$ $R(s)$-ratio measurements
show significant discrepancies between the ISR experiments KLOE and
BaBar, as well as between the scan experiments CMD-3 and SND20, for
which we still have no explanations, we learn that taking the $\tau$
data into account yields results that systematically increase $\amuh$,
making them compatible with the results of lattice QCD, in particular
with the BMW value.

The unexpected shift in theory has resulted from recent advances in
lattice calculations of the leading non-perturbative hadronic
contributions, the vacuum polarization \mbo{\amuh} and the light-light
scattering \mbo{\amuLbL}. In particular, the significantly higher
$\amuh$ result from BMW
2019~\cite{Budapest-Marseille-Wuppertal:2017okr,Borsanyi:2020mff}
and subsequent validations by the
Mainz/CLS~\cite{Ce:2022kxy,Kuberski:2024bcj,Djukanovic:2024cmq}
and RBC/UKQCD 2024~\cite{RBC:2018dos,RBC:2024fic}
collaborations have opened a new window for the $\amuh$
contributions. For the significant shifts resulting from the lattice
QCD results, the Muon g-2 Theory Initiative estimated in the WP25
update~\cite{Aliberti:2025beg}
\begin{equation}
\amuh=693.1(4.0) \to 713.2(6.1) \mbox{ \ and \ } \amuLbL=90(17) \to 112.6(9.6)
\label{thehadronicshifts}
\end{equation}
as necessary corrections. The resulting discrepancy between theory and
experiment has consequently decreased to
\begin{equation}
\delta a_\mu=a_\mu^{\rm exp}-a_\mu^{\rm SM}=38\pm 63 \times
10^{-11}
\end{equation}
and this deviation must be confronted with subdominant SM effects such
as the contribution of the weak interactions {12.9 $\sigma$}, the HLbL
effect 9.6 $\sigma$, and the higher order (HO) HVP effect {-8.2
  $\sigma$}. It is noteworthy that if one were to stick with the WP20
prediction, an apparently significant BSM effect $\Delta a_\mu^{\rm
  BSM} \stackrel{?}{=} \delta a_\mu$ would be 1.6 times larger than
the phenomenologically well-established weak
contribution. Furthermore, what we know from the intensive LHC
searches makes it very unlikely that a gap of 5.6 $\sigma$ could be
real. We now have the unbelievably precise agreement to 9 decimal
places in the muon $g$ factor.

The updated QED prediction of $a_\mu$ is
(see~\cite{Laporta:2017okg,Kurz:2016bau,Volkov:2017xaq,Aoyama:2024aly})
\begin{eqnarray}
a_\mu^{\rm QED}
 &=& \frac{\alpha}{ 2\pi}
+0.765,857,423(16) \left( \frac{\alpha }{ \pi}\right)^2
 +24.050,509,82(28) \left( \frac{\alpha }{ \pi}\right)^3
\nonumber\\
&& \hspace*{-1cm}+130.8734(60)\left( \frac{\alpha }{ \pi}\right)^4
+750.010(872)\left(\frac{\alpha }{ \pi}\right)^5
= 116584718.8(2) \power{-11}\epo
\end{eqnarray}
For the weak contributions, we have~\cite{Czarnecki:1995sz,Heinemeyer:2004yq,Gribouk:2005ee,Gnendiger:2013pva}
\begin{equation}
a_\mu^{\rm EW}=\underset{\text{LO}}{194.79(1)} \power{-11}  - \underset{\text{NLO}}{40.38(36)} \power{-11}=154.4(4) \power{-11}\epo
\end{equation}
The hadronic radiation corrections are the challenge for theory.
Contributions from hadrons (quark loops) at low energy scales are a
general problem in electroweak precision physics. About the muon
anomaly, three classes of hadronic contributions in which light quark
loops appear as hadronic ``blobs'' are: (a) Hadronic vacuum
polarization (HVP) of order $O(\alpha^2),O(\alpha^3)$; (b) Hadronic
light-light scattering (HLbL) of order $O(\alpha^3)$; (c) Hadronic
effects in 2-loop electroweak (EW) radiative corrections of order
$O(\alpha G_F m_\mu^2)$, $G_F$ the Fermi constant. The calculation
of non-perturbative effects is possible using hadron production data
in conjunction with the dispersion relations approach (DRA), effective
low-energy modeling within a resonance Lagrangian approach (RLA) and
ab initio lattice QCD calculations. Table~\ref{tab:hadew} provides a
summary of the hadronic and weak contributions (based on dispersive
approach),
\begin{figure}[t]
\centering
\includegraphics[width=0.495\textwidth]{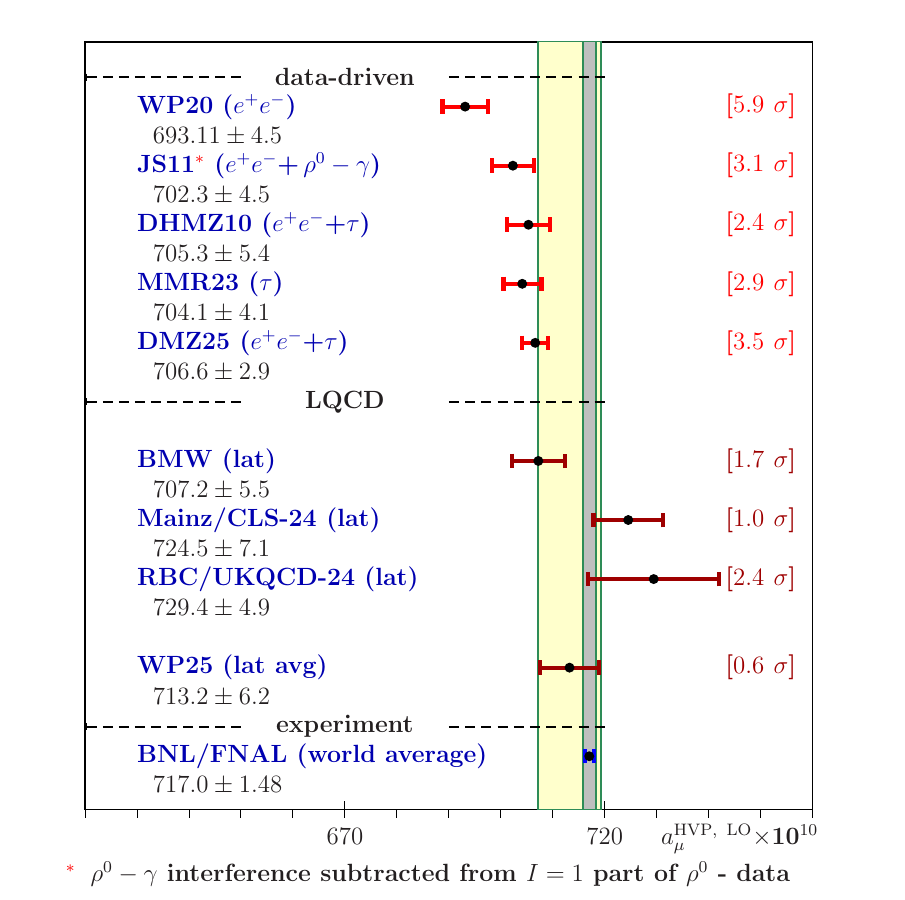}
\includegraphics[width=0.495\textwidth]{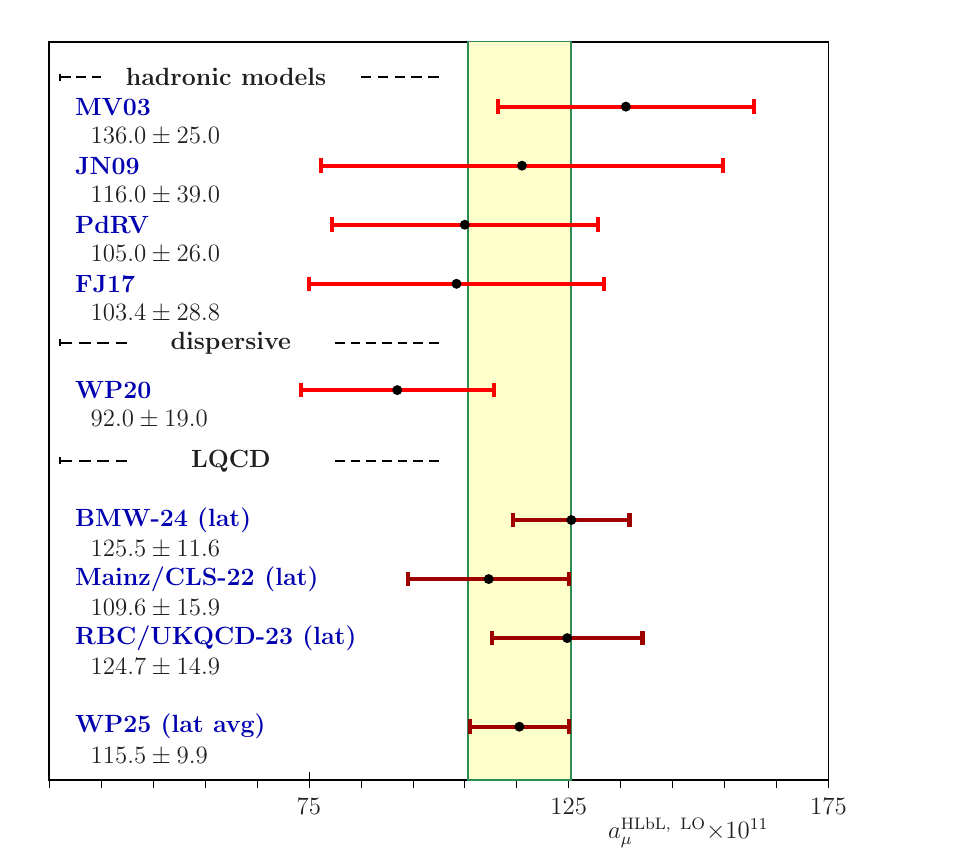}
\caption{Left: A selection of $\amuh$ results obtained using different
  approaches. WP20~\cite{Aoyama:2020ynm} is the dispersive result
  based exclusively on $\epm \to$ hadron data. JS11 refers to an
  $\epm$-data-based result corrected for $\rho^0-\gamma$ interference,
  which convincingly explains the puzzle surrounding the $\epm$-
  versus $\tau$-spectra relation~\cite{Jegerlehner:2011ti}. This
  correction is necessary to separate the irreducible QCD component
  from the normally ignored QED-QCD mixing inherent in the
  experimental $\pipi$ production data. DHMZ10~\cite{Davier:2010fmf}
  and the update DMZ25~\cite{Davier:2025jiq} have included the isospin
  breaking (IB)-corrected $\tau$ data in addition to the $\epm$
  data. MMR23 is an analysis based exclusively on $\tau$
  data~\cite{Miranda:2020wdg,Miranda:2020wdg}. BMW, Mainz/CLS, and
  RBC/UKQCD are the latest lattice QCD results, which currently
  provide the most reliable hadronic contributions. The last point is
  the update of WP25~\cite{Aliberti:2025beg} (wheat-colored band), the
  result of the consensus theory compared to the experimental result
  from BNL/FNAL (gray band), represented by the fictitious $\amuh$
  term required to close the gap between experiment and the theory
  prediction when dropping $\amuh$. Right: A selected history of
  $\amuLbL$ results. Pioneered by Hayakawa, Kinoshita and
  Sanda~\cite{Hayakawa:1995ps}, Bijnens, Pallante and
  Prades~\cite{Bijnens:1995cc} and Knecht and
  Nyffeler~\cite{Knecht:2001qf,Knecht:2001qg}. Data point shown are
  MV03~\cite{Melnikov:2003xd}, JN09~\cite{Jegerlehner:2009ry}, the
  consensus PdRV~\cite{Prades:2009tw}, my book
  FJ17~\cite{Jegerlehner:2017gek} and the result based on the
  dispersive approach by Colangelo et al.~\cite{Colangelo:2019uex} of the White Paper WP20. The
  lattice QCD results are BMW-24~\cite{Fodor:2024jyn},
  Mainz/CLS-22~\cite{Chao:2020kwq,Chao:2021tvp} and
  RBC/UKQCD-23~\cite{Blum:2017cer,Blum:2023vlm}. WP25 represents the
  current consensus of the Mon g-2 theory
  initiative~\cite{Aliberti:2025beg}.}
\label{fig:gm2staFJ25LQCD} 
\end{figure}
\begin{table}[h]
\centering
\caption{HVP, HLbL and weak contributions to $\amu$ where the HVP
  contributions rely on $\epm$ data. The HLbL is estimated mainly
  using transition form factors like $\gamma \gamma^* \to \pi^0$ and
  $\gamma \gamma^* \to \pi^+\pi^-,\pi^-\pi^0$.\\[-3mm]}
\label{tab:hadew}
\begin{tabular}{lcrl}
\hline\noalign{\smallskip}
$ a_\mu^{\mathrm{had}(1)}$&=&$(697.17\pm 4.18) \power{-10}$ &(LO) \\
$ a_\mu^{\mathrm{had}(2)}$&=&$(-10.89\pm 0.067) \power{-10}$ & (NLO)\\
$ a_\mu^{\mathrm{had}(3)}$&=&$(1.242\pm 0.010) \power{-10}$ &
(NNLO)\\
$ a_\mu^\mathrm{had, LbL}$&=&$(10.34\pm 2.88) \power{-10}$ & (HLbL)\\
$ a_\mu^{\mathrm{weak}}  $&=&$(15.4\pm 0.4) \power{-10}$ & (LO+NLO)\epo\\
\noalign{\smallskip}\hline\\[-5mm]
\end{tabular}
\end{table}
\noindent where the LO-HVP and LO-HLbL results (see
Fig.~\ref{fig:gm2staFJ25LQCD}) have to be updated according to
(\ref{thehadronicshifts}) by LQCD results!  The uncertainties of the
subleading NLO and NNLO and weak contributions are below one $\sigma$,
i.e., they are ``DRA save''.
\section{What is the problem?}
Here we have to remind the role of $\tau$ decay spectra for the
$\amuh$ evaluation. The lesson we learn from
Fig.~\ref{fig:gm2staFJ25LQCD} is that including the charged channel
$\tau \to \nu_\tau \pi \pi^0 $ spectra brings the data-driven
dispersive results close to the lattice QCD results from the
BMW Collaboration. The problem with
the DRA-HVP evaluation must therefore be related to the experimental
$\epm$ data.

Despite determined efforts to calculate and model QED corrections for
hadronic processes, open questions remain. What can explain the large
difference between experimental $\epm$-HVP data and lattice QCD-HVP
data? In the process \mbo{e^+e^- \to {\rm hadrons}}, experiments
measure the photon propagator, i.e., a mixture of QCD and QED effects
that is difficult to disentangle. This requires the removal of
external QED corrections, where the radiation of hadrons is not well
understood, even though scalar QED (sQED) is a good approximation for
meson production at low energies. In contrast, lattice QCD measures the
hadronic blob exclusively, as a correlator of two hadronic currents:\\[-6mm]
\begin{figure}[h!]
\centering
\includegraphics[height=1.5cm]{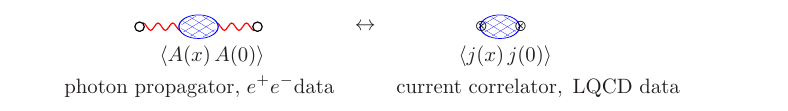}
\end{figure}

\vspace*{-3mm}

\noindent
It is interesting to compare the dipion spectra of $\epm \to \pipi$
and \mbo{\tau^- \to \nu_\tau \pi^- \pi^0} taking into account the
Isospin Breaking (IB) effects. In many respects, the $\tau$ spectrum
is much simpler than that of \mbo{\epm}, since the charged
\mbo{\rho^\pm} does not mix with other states in the dominant
low-energy range below about 1.05 GeV (above \mbo{\phi}, below
\mbo{\rho'}). The $\tau$ decays are mediated by the very heavy $W$
boson, in stark contrast to $\epm$ annihilation, which is mediated far
below the $Z$ boson by the massless photon, which mixes with
$\rho^0,\omega$, and $\varphi$.  $\tau$ spectra provide isovector
$I=1$ dipion spectra which should agree with the $I=1$ part of the
$\epm$ annihilation upon appropriate isospin breaking corrections.  In
contrast to the charged current (CC) data, understanding the neutral
current (NC) data requires more complex QCD-driven hadron
phenomenology modeling: Chiral perturbation theory (CHPT) has been
extended to include spin-1 resonances, which are well described by the
vector meson dominance model (VMD) or, in perfect form, by the
Resonance Lagrangian Approach (RLA). These models predict the dynamic
widths and dynamic mixing of $\gamma, \rho^0, \omega, \phi$ rather
accurately.

A key effect that is often overlooked is the mixing of $\rho^0$,
$\omega$, and $ \varphi$ with the photon, especially the
$\rho^0-\gamma$ mixing, which directly impacts the relationship
between the photon propagator and the One-Particle-Irreducible (1PI)
HVP blob. This mixing appears in the NC measurements \mbo{\epm \to
  \pi^+\pi^-}, but not in the CC data \mbo{\tau^- \to \nu_\tau \pi^-
  \pi^0}. This issue was examined in~\cite{Jegerlehner:2011ti} to
clarify the known discrepancy between the dipion spectra from $\tau$
decays $\varv_\pm(s)$ and the $\epm$ data $\varv_0(s)$, which
persisted despite applying the commonly accepted isospin-breaking
correction (IB) $R_{\rm IB}(s)$ to the $\tau$ data.

\begin{figure}
\centering
\includegraphics[width=0.48\textwidth]{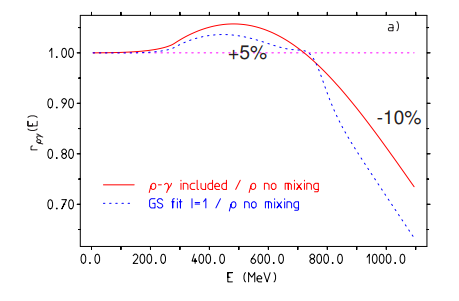}
\includegraphics[width=0.48\textwidth]{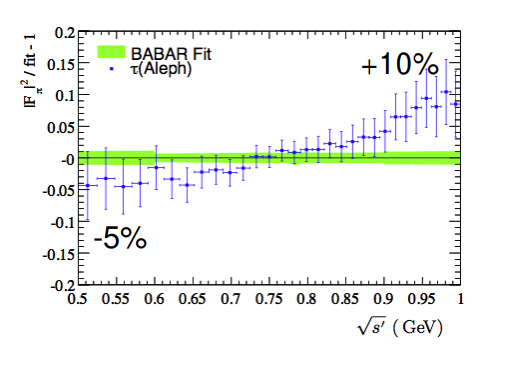}
\caption{Left: The ratio of the isospin $I=1$ pion form factor
  $|F_\pi(E)|^2$ taking mixing into account, normalized to the case
  without mixing. Also shown is the same ratio of the $I=1$ part of
  the $\epm$ data to the $\tau$ data GS fits, mimicked by fictitious
  parameter shifts in mass and width. Right: Best ``proof'' for our
  $\gamma-\rho^0$ mixing profile is the ratio of the ALEPH $\tau$
  decay spectrum versus the BaBar $\epm$ spectrum~[reproduced as part
    of Fig.~55 in {\tt arXiv:1205.2228} by J.~P.~Lees et al.]~\cite{BaBar:2012bdw}] (also
see~\cite{Davier:2010fmf}).}
\label{fig:mixingcorr}
\end{figure}

In fact, taking into account the \mbo{\rho-\gamma} interference solves
the mystery of \mbo{\tau} (charged channel) vs \mbo{e^+e^-} (neutral
channel). The \mbo{\rho-\gamma} interference (which does not occur in
the charged channel) is often mimicked by large shifts in the mass
$M_\rho$ and width $\Gamma_\rho$, but such large shifts are not
consistent with known calculations of $M_{\rho^\pm}-M_{\rho^0}$, for
example according to the Cottingham formula. The missing element in
the standard derivations of the Gounaris-Sakurai formula (based on the
VMD-I model, which breaks electromagnetic gauge invariance) is the
omission of the $\gamma - \rho^0$ mixing propagator, which we
calculated using the gauge-invariant VMD-II approach together with
sQED, which should be valid below 1 GeV (as can be
learned from the $\gamma\gamma \to \pi^+\pi^- \mbox{ \ vs \ } \pi^0
\pi^0$ data). We thus are confronted with a non-diagonal
$(\gamma,\rho^0)$ $2\,\times\,2$ matrix propagator, which we must
diagonalize. We do not obtain the correct diagonalization by omitting
the non-diagonal mixing term. The $\gamma-\rho^0$ mixing effect, which
is proportional to $eg$ in terms of couplings, where $e$ is the
electric charge and $g$ is the effective coupling $\rho \pi\pi$,
undergoes a resonance amplification, resulting in unexpectedly large
distortions of the I=1 resonance peak\footnote{This correction is
robust, as it involves no new parameters that do not already appear in
the GS formula. The effect only depends on the well-known leptonic
$\rho$ width $\Gamma_{\rho \to ee}$.}, as shown in
Fig.~\ref{fig:mixingcorr}.

In~\cite{Jegerlehner:2011ti}, we \textbf{wrongly} corrected the $\tau$
spectral function\footnote{ That the correction $r_{\rho\gamma}(s)$
has to be applied to the $\epm$ data was already stated and discussed
in Sect.~21.3 of~\cite{Benayoun:2019zwh}.}
$\varv_-(s)$ as follows

\be
\varv_0(s)=r_{\rho\gamma}(s)\,R_{\rm IB}(s)\,\varv_-(s) \epo \nn
\ee
But as we just argued, it is not the \mbo{\tau} spectra that require correction
for the absence of \mbo{\rho-\gamma} mixing. Instead, this QED-QCD
interference effect must be removed from the \mbo{e^+e^-} data in order to get
the purely hadronic part of the 1PI
self-energy. Thus the proper correction must read:
\begin{equation}
\left.\varv_0(s)\right|_{\rm
  corr}=\varv_0(s)/r_{\rho\gamma}(s)=R_{\rm IB}(s)\,\varv_-(s)\epo
\end{equation}
In fact, as in the $\tau$-channel, the $\gamma-\rho^0$ mixing is
missing in the LQCD data, since this data are obtained by simulating
the ``QCD-only'' path integral\footnote{Intrinsic QED
effects relevant for HVP, such as final state radiation (FSR), must be
included separately.}.

For the $\rho^0$ resonance range [0.63,0.96] GeV, the $\rho-\gamma$
interference leads to a shift~\cite{Jegerlehner:2011ti}\footnote{It is
important to apply the correction only to the $I=1$ contribution. Due
to the narrow widths of $\omega$ and $\varphi$, the mixing for these
states is much smaller.}.
\be
\delta \amuh [\rho\gamma]\simeq (5.1\pm0.5)\power{-10}\epo
\ee
This correction must be added to the
standard \mbo{e^+e^-}-based \mbo{a_\mu^{\rm HVP,\ LO}}.
\be
a_\mu^{\rm HVP,\ LO}\left|_{\rm
  corr}\right.=a_\mu^{\rm HVP, \ LO}+\delta \amuh [\rho\gamma]\simeq
(702.3 \pm 4.2)\power{-10}\epo
\label{eecorr}
\ee
This result fits well with $\amuh [ee+\tau]=(705.3\pm 4.5)\power{-10}$
from Davier et al.~\cite{Davier:2010fmf}, where \mbo{\tau} decay
spectra are taken into account.

In summary, removing the $\rho^0-\gamma$ mixing from the $\epm$ data
is essential, as it brings the dispersive result into better agreement
with the BMW lattice result, $\amuh = (707.5 \pm 5.5) \power{-10}$.

With respect to the value $\amuh=713 \power{-10}$ accepted by WP25, we
conclude that at least part of the missing shift is explained. If the
BMW results were to prevail over the WP25 value, the $\rho^0-\gamma$
mixing would largely explain the discrepancy between the dispersive
result based on $\epm$ data and the lattice and $\tau$-based results.

\section{Electron g-2 Status and its Future}  
Here is a brief note on the status of the electron magnetic moment
$a_e$, which is a test of QED that is highly sensitive to $\alpha$
electromagnetic as determined by atomic interferometry. In 2018, with
$\alpha^{-1}({\rm Cs18})=137.035999046 (27)$ one had $a_e^{\rm
  exp}-a_e^{\rm the}=(-84\pm36)\power{-14}$, a deviation of
$-$~2.3~$\sigma$ between theory and experiment. This presented a
difficult situation for developers of BSM models, as the sign of
$\delta a_e$ differed from that of $\delta a_\mu$, which was difficult
to explain.  In 2020, however, a more precise value of
$\alpha^{-1}({\rm Rb20})=137.035999206 (11)$ was
obtained~\cite{Morel:2020dww}, which meant that $a_e^{\rm
  exp}-a_e^{\rm the}=(51\pm30)\power{-14}$ changed sign and now
predicted a gap of +~1.7~$\sigma$. Note that $\alpha$, the most
fundamental parameter in physics, changed by 5.4~$\sigma$~ due to the
switch from Cesium (Cs) to Rubidium (Rb) atoms! Soon after, in 2022, a
more accurate measurement improved the value of $a_e^{\rm
  exp}=0.00115965218073(28)$ to $a_e^{\rm exp}=0.00115965218059
(13)$~\cite{Fan:2022eto}. But the QED prediction was also improved. In
2024, the universal 5-loop coefficient changed to
\mbo{A_1^{(10)}=5.873(128)}~\cite{Volkov:2017xaq} (crosschecked
in~\cite{Aoyama:2024aly}), which then leads to the
prediction\footnote{In this result, also $a_e^{\rm LO-HVP}$ had to be
slightly modified to take into account the change in $a_\mu^{\rm
  LO-HVP}$ (lattice vs dispersive result), i.e., $a_e^{\rm
  LO-HVP}=1.871(11) \times 10^{-12}$ changes to $a_e^{\rm
  LO-HVP}=1.923(09) \times 10^{-12}$.}
\begin{equation}
a_e^{\rm the}=0.00115965218023(9) \mbox{ \ and \ } a_e^{\rm
  exp}-a_e^{\rm the}=(36\pm16)\power{-14}\epo
\end{equation}
Only a deviation of +~2.3~$\sigma$, which still agrees quite well with
the SM prediction. In future atomic interferometry experiments (as
part of the AION project ~\cite{Badurina:2019hst}) based on
Strontium and Yttrium, are expected to reduce the uncertainty of the
electromagnetic fine structure constant such that an $a_e$ prediction
with an accuracy of $2\power{-14}$ would be possible.
\section{Conclusions}
One of the most famous oracles that promised new physics beyond the SM
has dissolved, thereby consolidating the SM up to the TEV scale, which
is consistent with the LHC searches! However, this does not mean that
the search for BSM physics as the main goal of particle physics is
losing importance. What do $a_\mu$ and $a_e$ tell us?  High-precision
physics may prove more difficult than expected to reach the limits of
the SM, and this applies to both $a_\mu^{\rm exp}$ and $a_\mu^{\rm
  the}$. For the electron, the limitation arises from the need for
extremely precise atomic interferometry to determine the
electromagnetic fine structure constant in the Thomson limit. As far
as $a_\mu^{\rm exp}$ is concerned, all experiments to date (CERN, BNL,
FNAL) have used the magic $\gamma$ trick, and an alternative type of
experiment is urgently needed. Hopefully, the J-PARC
project~\cite{Saito:2012zz,Mibe:2011zz,Venanzoni:2025rsd}, which is
promoting an experiment with ultra-cold muons in an $\vec{E}=0$ field
cavity, will be realized in the not-too-distant future.

Given the failure to provide a clear answer based on data-driven DRA,
the MUonE determination (elastic $\mu e$ scattering) of the HVP
at CERN (directly measuring what LQCD calculates) would be a mandatory
experiment that would need to be carried
out~\cite{CarloniCalame:2015obs,MUonE:2016hru,Spedicato:2025ofo,Price:2025pru}.
Lattice QCD has proven to be an indispensable tool for predicting
non-perturbative hadronic effects in electroweak precision
observables. However, since a significant fraction of the results
still have to be estimated by extrapolation to small lattice spacings
and large volumes, further progress is required to reduce the
extrapolation uncertainties.

As for an improved $\alpha$ determination as input for predicting
$a_e$, the AION project (using Strontium and Yttrium
ions)~\cite{Badurina:2019hst} promises significant progress.

Although lattice QCD already provides satisfactory results for the
hadronic vacuum polarization, it remains an urgent task to understand
the inconsistencies in the experimental data required in the
dispersive approach: KLOE vs BaBar, SND20 vs CMD-3, $\epm \to
\pi^+\pi^-$ vs $\tau^- \to \pi^0 \pi^- \nu_\tau$ data (NC vs CC); A
better determination of the HLbL contribution requires more
$\gamma\gamma \to {\rm hadrons}$ data to improve the precision of
known channels and explore new channels that have not been
experimentally accessible so far.

Beyond BSM science fiction: Scrutinizing SM predictions and making
progress in determining SM parameters such as $M_t$ and $M_H$ (which
requires a high-precision Higgs/top quark pair factory) is a major
challenge that is also important for a better understanding of early
cosmology. But that would be another topic~\cite{Jegerlehner:2021vqz}!

Outlook: We note that we have remarkable agreement between the
following results:\\[3mm]
\begin{tabular}{lccr}
 $a_\mu ^{\rm HVP-LO}[ee+\gamma\rho]$& = &$702.3 \pm 4.2\power{-10}$ & Eq.~(\ref{eecorr}) \\
 $a_\mu^{\rm HVP-LO}[ee+\tau]$& = &$705.3\pm 4.5\power{-10}$ &\cite{Davier:2010fmf} \\
  $a_\mu^{\rm HVP-LO}[\tau]$& = &$704.1 \pm 4.1 \times 10^{-10}$ & \cite{Masjuan:2023yam} \\
 $a_\mu^{\rm HVP-LO}[\rm LQCD:BMW]$& = &$707.5\pm 5.5 \times 10^{-10}$ &\cite{Borsanyi:2020mff}\\[3mm]
\end{tabular}

\noindent
Since Mainz/CLS and RB/UKQCD achieved slightly larger lattice results
(compared to BMW), the 2025 White Paper combined the lattice results
to give the larger value $a_\mu ^{\rm HVP-LO}=713.2 \pm 6.2
\power{-10}$.  If the progress in the lattice results confirmed the
slightly smaller BMW value, the mystery surrounding the discrepancy
between the dispersive approach and the lattice results would be
largely solved if the $\epm$ data were corrected for the $\rho^0 -
\gamma$ interference. Then, the difference between the theoretical
prediction and the experimental measurement at a level of 0.9 $\sigma$
would still provide strong confirmation of the SM. In fact, the hybrid
evaluation, which combines the “best” of the lattice and dispersive
approaches:~\cite{Davier:2023cyp,Boccaletti:2024guq} [BMW/DMZ-24],
highlighted the result as “confirmation of the SM up to a level of
0.37 ppm.”

\noindent
{\bf Acknowledgments} \\ Many thanks to the organizers for the kind
invitation to the “Matter to the Deepest” Conference 2025 in Katowice
and for allowing me to give this talk. I would also like to express my
sincere gratitude for your kind support and the inspiring atmosphere
at the Conference.

\end{document}